\documentclass[preprint,showpacs,preprintnumbers,amsmath,amssymb]{revtex4}

\def\beq{\begin{equation}}
\def\eeq{\end{equation}}
\newcommand{\elkobar}[1]{\overset{{}^{{}^{\boldsymbol{\neg}}}}{\smash[t]{#1}}}
\usepackage{graphicx}% Include figure files
\usepackage{dcolumn}% Align table columns on decimal point
\usepackage{bm}% bold math

\begin{document}

%\preprint{APS/123-QED}

\title{MiniBooNE and a $\mathbf{(CP)^2 = -\openone} $ sterile neutrino} 
% Force line breaks with \\

\author{D. V. Ahluwalia-Khalilova}
\email{dharamvir.ahluwalia-khalilova@canterbury.ac.nz}
\author{Alex B. Nielsen}
\email{abn16@student.canterbury.ac.nz}
\affiliation{%
 Department of Physics and Astronomy, Rutherford Building, University of Canterbury, Private Bag 4800, Christchurch 8020, New Zealand}

\date{16 February 2007, Accepted for publication in Mod. Phys. Lett. A
}% It is always \today, today,

\begin{abstract}
It has been taken as granted that the 
observation of two independent mass-squared
differences necessarily fixes the number of underlying mass eigenstates
as three, and that the addition of a sterile neutrino provides an additional
mass-squared difference. The purpose of this Letter is to argue that
if one considers
a sterile neutrino component that belongs to the $(CP)^2 = -\openone$  
sector, then both of the stated claims are false. We also
outline how the results reported here, when combined with
the forthcoming MiniBooNE data and other
experiments, can help settle the issue of the $CP$ properties of the sterile
neutrino; if such a component does indeed exist.
\end{abstract}

\maketitle

To understand the excess $\overline{\nu}_e$ events at
LSND~\cite{Athanassopoulos:1996jb}, and their absence at
KARMEN~\cite{Armbruster:1998uk}, is one of the outstanding issues of
neutrino-oscillation phenomenology. On the experimental side, the
forthcoming results from MiniBooNE~\cite{Stancu:2006gv,Ray:2007yd}
should shed significant light on this issue.  On the theoretical side,
it has been generally accepted that the observation of two independent
mass-squared differences necessarily fixes the number of underlying
mass eigenstates to three, and that the addition of a sterile neutrino
provides an extra mass-squared difference. The purpose of this Letter
is to argue that if one considers a sterile neutrino component
belonging to the $(CP)^2 = -\openone$ Wigner sector~\cite{Wigner1962},
then both of the stated claims are false. We also outline how the
existing data, and the forthcoming results from MiniBooNE 
and other experiments, can help
settle the issue of the $CP$ properties of the sterile neutrino; if
such a component does indeed exist.

That $(CP)^2= -\openone$ classes exist for spin one half, and that
these remain fermionic, is a classic result obtained by Wigner in the
early sixties~\cite{Wigner1962}, and later by
Lounesto~\cite{Lounesto1997}. This happens without going beyond the
Poincar\'e symmetries.  Recently, one such quantum field has been
explicitly
constructed~\cite{Ahluwalia-Khalilova:2004sz,Ahluwalia-Khalilova:2004ab},
and its various properties have been investigated by a number of
authors~\cite{daRocha:2005ti,Boehmer:2006qq,Boehmer:2007dh}.  For
instance, the explicit construct has been shown to carry limited
interactions with standard model fields.  In the absence of a
preferred direction \textemdash~such as the one which may arise due to
an external magnetic field \textemdash~it carries a Klein-Gordon
propagator.  Here, however, we do not restrict ourselves to the
$(CP)^2 = -\openone$ states reported in
~\cite{Ahluwalia-Khalilova:2004sz,Ahluwalia-Khalilova:2004ab} but
allow for a more general identification. The basic inspiration to take
from all this is that Dirac, or Majorana, eigenstates do not exhaust
the possibilities that exist for a spin one half fermionic field, but
that additional, equally fundamental, constructs exist and that these
carry new and unexpected physical properties. These should be
experimentally investigated without theoretical prejudices.  Neutrino
oscillations, as will become apparent in what follows, provide an
ideal laboratory to probe these aspects of physical reality.

Towards the stated goal, as an ansatz, we consider a sterile neutrino
that belongs to $(CP)^2 = -\openone$ sector of the spin-$\frac{1}{2}$
fermionic Wigner classes~\cite{Wigner1962}, and define a set of four
flavour neutrino states~\footnote{Such a suggestion has already been
noted by Bernardini in Ref.~\cite{Bernardini:2006cn}.}
\beq
\vert \xi \rangle =  \sum_{j=1,2,3} 
                     U_{\xi j}\vert d_j, h \rangle + U_{\xi 4}
\vert 4, h \rangle \label{eq:fi} 
\eeq
where the flavour $\xi=\alpha,\beta,\gamma,\delta$. The $U_{\xi j}$
and $U_{\xi 4}$ are elements of a $4\times 4$ real unitary matrix
carrying no CP violation. This assumption is merely for simplicity and
may be relaxed \footnote{By adding the $(CP)^2 = \pm \openone$ sectors
together, however, we do allow for the possibility of CP
violation. This may lead to mixing of the $(CPT)^2 = \pm \openone$
Wigner classes and to a novel form of $CPT$ violation. The latter is
presently widely considered in the context of neutrino-oscillation
data, see, e.g., Ref.~\cite{Murayama:2000hm,Ahluwalia:1998xb}.  For
additional discussion of this point, in a related context, the reader
is referred to
Ref.~\cite{Ahluwalia-Khalilova:2004sz,Ahluwalia-Khalilova:2004ab}}.
The $\vert d,h \rangle$ are Dirac mass eigenstates, with $(CP)^2 = +
\openone$; while $\vert 4,h \rangle$ is a sterile mass eigenstate,
with $(CP)^2 = - \openone$. Each of the four mass eigenstates carries
a different mass, and $h$ denotes helicity. The assumption that
$(CP)^2=-\openone$ for the sterile component guarantees that it is not
a Dirac mass eigenstate.  Apart from the stated $(CP)^2$ properties,
the introduced Dirac and sterile sectors carry the following
additional $CP$ properties

\vspace{5pt}
\noindent
$\mathbf{Dirac},\;\{C,P\}=0:$ 
\begin{eqnarray}
CP \vert d, h\rangle = \vert\bar{d},-h\rangle,\; 
CP \vert \bar{d}, h\rangle = \vert d,-h\rangle.       
\end{eqnarray}
$ \mathbf{Sterile},\;[C,P]=0:$
\begin{eqnarray}
&& CP \vert 4, h \rangle  = - i \vert \elkobar{4}, -h\rangle,\;
  CP \vert 4, - h \rangle  = + i \vert \elkobar{4}, h\rangle \\
&& CP \vert \elkobar{4}, h \rangle  = - i \vert 4, -h\rangle,\;
  CP \vert \elkobar{4}, - h \rangle  = + i \vert 4, h\rangle.
\end{eqnarray}

The former are in accord with the careful treatment given in
Ref.~\cite{Weinberg1995}. The latter, after appropriate
re-identifications, coincide with the $(CP)^2=-\openone$ construct of
Ref.~\cite{Ahluwalia-Khalilova:2004sz,Ahluwalia-Khalilova:2004ab};
where, incidentally, it corresponds to the $(CPT)^2=-\openone $ Wigner
class. For the chosen sterile sector, $CP$ is an anti-unitary
operator. The $\vert\bar d\rangle$ represents the CP conjugate of a
Dirac mass eigenstate $\vert d\rangle$; while generically
$\vert\elkobar{\lambda}\rangle $, with $\lambda=4$ above, denotes the CP
conjugate of a mass eigenstate with $(CP)^2 = -\openone$.

Using the above enumerated results, the successive action of $CP$ on
the set of flavour eigenstates (\ref{eq:fi}) yields the following flavour
states

\begin{eqnarray}
&&\hspace{-21pt} \vert \tilde{\xi} \rangle := CP  \vert \xi \rangle
=
 \sum_{j} U_{\xi j}\vert \bar{d}_j, - h \rangle  - i U_{\xi 4}
\vert \elkobar{4}, - h \rangle \label{eq:fi2}\\
&&\hspace{-21pt} \vert \xi^\prime \rangle := CP \vert \tilde{\xi} \rangle 
=  \sum_{j} 
                     U_{\xi j}\vert d_j, h \rangle - U_{\xi 4}\vert 4, 
h \rangle \label{eq:fi3} \\
&& \hspace{-21pt} 
\vert \tilde{\xi}^\prime \rangle := CP  \vert \xi^\prime 
\rangle
=
\sum_{j} U_{\xi j}\vert \bar{d}_j, - h \rangle  + i U_{\xi 4}
\vert \elkobar{4}, - h \rangle \label{eq:fi4}
\end{eqnarray} 
with $CP \vert \tilde{\xi}^\prime \rangle$ being identical to the
original flavour $\vert \xi \rangle$ defined in Eq.~(\ref{eq:fi}).
Here, and in the following, 
$
j,k:= 
1,2,3 
$
corresponds to the three Dirac mass eigenstates. 
The flavour index $\xi$, as already mentioned, runs over four flavours
\beq
\xi:= \alpha,\beta,\gamma,\delta \label{eq:xiz}.
\eeq 
These flavours would soon be seen to
be connected to, but not identical with, the 
operationally defined flavours of neutrinos.
The circumstance that

\beq
\vert \xi \rangle \stackrel{CP}{\rightarrow}
\vert \tilde{\xi} \rangle  \stackrel{CP}{\rightarrow}
\vert \xi^\prime \rangle \stackrel{CP}{\rightarrow}
\vert \tilde{\xi}^\prime \rangle \stackrel{CP}{\rightarrow} \vert \xi \rangle
\label{eq:4cp}
\eeq
forces upon us the fundamental question: 
\begin{quote}
How are the above-obtained flavour states related to the flavour
eigenstates $\vert \nu_\ell\rangle$ and $\vert\bar\nu_\ell\rangle$,
where $\ell$ corresponds to the operationally-defined flavours
$\ell:=e,\mu,\tau,\zeta$
\footnote{The new flavour $\zeta$  
is just a logical necessity of the $4\times 4$ framework. It
complements the three flavours associated with electron, muon, and tau
charged leptons.}?

\end{quote}
To answer this question we first re-write (\ref{eq:4cp}) by collecting
together flavours connected by $(CP)^2$ \footnote{In the absence of
the considered sterile component the flavours connected by $(CP)^2$
are identical. Now this is no longer so.}.  This gives rise to a pair
of flavour sets
\beq
\left\{
\begin{array}{ll}
\vert \xi \rangle \\
\vert \xi^\prime \rangle
\end{array}
\right\},\quad
\left\{
\begin{array}{ll}
\vert \tilde\xi \rangle \\
\vert \tilde\xi^\prime \rangle
\end{array}  
\right\}\label{eq:fset}.   
\eeq
The action of $CP$ is now encoded in the following schematic equation
\beq
CP \left\{
\begin{array}{ll}
\vert \xi \rangle\\
\vert \xi^\prime \rangle
\end{array}  
\right\}\rightarrow
\left\{
\begin{array}{ll}
\vert \tilde\xi \rangle\\
\vert \tilde\xi^\prime \rangle
\end{array}  
\right\},\;\;
CP \left\{
\begin{array}{ll}
\vert \tilde\xi \rangle\\
\vert \tilde\xi^\prime \rangle
\end{array}  
\right\}\rightarrow
\left\{
\begin{array}{ll}
\vert \xi^\prime\rangle\\
\vert \xi \rangle
\end{array}  
\right\}.
\eeq
The action of $CP$ thus rotates between the two flavour sets.
This circumstance leads us to suggest the following
identification

\begin{quote}
In any given production of $\nu_\ell$ one either creates a
$\vert\xi\rangle$ or $\vert\xi^\prime\rangle$ with equal
probability. Similarly, $\bar\nu_\ell$ creation corresponds to the
production of either $\vert\tilde\xi\rangle$ or
$\vert\tilde\xi^\prime\rangle$ with equal probability.

\end{quote}
So, for example,
\beq
\bar\nu_\mu:= \left\{
\begin{array}{ll}
\vert \tilde\beta \rangle \\
\vert \tilde\beta^\prime \rangle
\end{array}  
\right\}, \quad
\bar\nu_e:= \left\{
\begin{array}{ll}
\vert \tilde\alpha \rangle \\
\vert \tilde\alpha^\prime \rangle\label{eq:fdef}
\end{array}  
\right\}.
\eeq
The discussion above constitutes our working answer to the asked
question. At this stage it is an hypothesis. Its validity, or its
final acceptance, should be deferred to experiments, and/or to
additional theoretical work. What appears certain is that if nature
does superimpose mass eigenstates with differing $(CP)^2$, or for that
matter with different $(CPT)^2$ properties, then the
particle-antiparticle concept must undergo a fundamental
reexamination. Our suggestion constitutes a preliminary, and perhaps
first, theoretical attempt in that direction. However, its flavour
mirrors the remarks contained in the opening
paragraph of Sec. III A of Langacker-London paper on nonorthogonal
neutrinos~\cite{Langacker:1988up}.\footnote{The idea of non-orthogonal neutrinos
seems to have first arisen in Ref.~\cite{Lee:1977qz}.}

One of our tasks here is to understand 
excess $\overline{\nu}_e$ events at
LSND~\cite{Athanassopoulos:1996jb}, their absence at
KARMEN~\cite{Armbruster:1998uk}, and the forthcoming results from the
MiniBooNE~\cite{Stancu:2006gv,Ray:2007yd}.
As such, we now exploit the emergent interpretation for calculating 
the flavour oscillation probability for $\bar\nu_\mu \to \bar\nu_e$.
It is given by
\begin{eqnarray}
\hspace{-21pt}\mathcal{P}\left(\bar\nu_\mu\to\bar\nu_e\right)&=&
\frac{1}{4}\bigg\{
\Big\vert\left\langle\tilde\alpha\Big\vert\tilde\beta\right
\rangle^t\Big\vert^2
+\Big\vert\left\langle\tilde\alpha^\prime\Big\vert\tilde\beta
\right\rangle^t\Big\vert^2 \nonumber\\
&&\hspace{11pt} +\Big\vert\left\langle\tilde
\alpha\Big\vert\tilde\beta^\prime\right\rangle^t\Big\vert^2
+\Big\vert\left\langle\tilde\alpha^\prime\Big\vert\tilde\beta^\prime
\right\rangle^t\Big\vert^2
 \bigg\}\label{eq:pmbeb}
\end{eqnarray}
where the notation $\vert ~\rangle^t$ corresponds to the space-time evolved
state in the usual neutrino-oscillation experimental setting.
In order to understand the ensuing expression for the 
$\mathcal{P}\left(\bar\nu_\mu\to\bar\nu_e\right)$, we would now like
to present each of the four terms explicitly. To facilitate this, we 
introduce the following definitions

\begin{eqnarray}
&&\mathcal{A}_{jk} := 4 U_{\alpha j} U_{\alpha k} U_{\beta j} U_{\beta k}
\label{eq:ajk}   \\
&& \mathcal{\chi} := \frac{1}{2}\bigg\{\sum_{j} 
U_{\alpha j} U_{\beta j} - U_{\alpha 4} U_{\beta 4} \bigg\}^2 
\label{eq:apm}\\
&& \varphi_{jk}:=\frac{\Delta m_{jk}^2 L}{4 \hbar c E} = \frac{1.27  
\Delta m_{jk}^2 
(\mathrm{eV^2}) L (\mathrm{km})}{E(\mathrm{GeV})}  \label{eq:dmjk} 
\end{eqnarray}  
where $L$ refers to the source-detector distance, $E$ is the neutrino energy,
and $\Delta m_{jk}^2:= m_j^2-m_i^2$. 
We also need to define $\mathcal{A}_{j4}$ and  
$\Delta m_{j4}^2$. These are obtained by the replacement
$k \to 4$ in Eq. (\ref{eq:ajk}) and  Eq. (\ref{eq:dmjk}),
respectively.

With these definitions, the  four terms 
that appear in Eq.~(\ref{eq:pmbeb}) read
\begin{eqnarray}
&& \hspace{-21pt} \Big\vert\left\langle\tilde\alpha\Big\vert\tilde\beta\right
\rangle^t\Big\vert^2 = \Big\vert\left\langle\tilde\alpha^\prime
\Big\vert\tilde\beta^\prime\right\rangle^t\Big\vert^2 
\nonumber \\ 
&& =
 - \sum_{j<k}\mathcal{A}_{jk}\sin^2 \varphi_{jk}
 -\sum_{j} A_{j 4} \sin^2 \varphi_{j4} \nonumber
\\
&&\hspace{-21pt} \Big\vert\left\langle\tilde\alpha^\prime\Big\vert\tilde\beta
\right\rangle^t\Big\vert^2 = 
\hspace{11pt} \Big\vert\left\langle\tilde
\alpha\Big\vert\tilde\beta^\prime\right\rangle^t\Big\vert^2 
\nonumber \\ 
&& =
2 \chi
 - \sum_{j<k}\mathcal{A}_{jk}\sin^2 \varphi_{jk}
 +\sum_{j} A_{j 4} \sin^2\varphi_{j4}.\nonumber
\end{eqnarray}
As such, we have the central result of the Letter 

\beq
\mathcal{P}\left(\bar\nu_\mu\to\bar\nu_e\right)
= \chi
 - \sum_{j< k}\mathcal{A}_{jk}\sin^2 \varphi_{jk}.
\eeq
It is to be immediately noted that due to the manifest cancellation of
the $\varphi_{j4}$ terms, the $(CP)^2 = -\openone$ sterile-neutrino
component does \textit{not} induce any oscillatory terms.  Its
presence is felt solely through the constant \textit{flavour
transmutation} term.  The transmutation term may be re-written as
\beq 
\chi =
 \bigg\{ \sum_{j} 
U_{\alpha j} U_{\beta j} \bigg\}^2 + \big\{U_{\alpha 4} U_{\beta 4}\big\}^2  
\label{eq:ft}.
\eeq
It represents a fundamentally unavoidable $\bar\nu_e$
``contamination'' in a $\bar\nu_\mu$ beam. 
We refer to this
``contamination'' as a flavour transmutation rather than a
flavour oscillation due to its $L/E$ independence.
Its origin can be traced back to the fact that 
while each of the flavours,
$\alpha,\beta,\gamma,\delta$ in (\ref{eq:fi2})
are mutually orthogonal (e.g., $\langle\tilde\alpha\vert\tilde\beta\rangle = 0$),
 and the same being true with the flavours, say, in
(\ref{eq:fi4}) (e.g., $\langle\tilde\alpha^\prime\vert\tilde\beta^\prime\rangle = 0$);
the 
flavour states
such as $\vert \tilde\alpha\rangle$   and 
$\vert \tilde\beta^\prime \rangle$ 
have non-vanishing overlap, i.e. $\langle\tilde\alpha\vert\tilde \beta^\prime 
\rangle \ne 0$.

The above remarks parallel those surrounding Eq.~(25) of 
the Langacker-London paper on nonorthogonal
neutrinos~\cite{Langacker:1988up}. There, the effect
 arises from a mismatch between the light and heavy neutrinos. 
In our case a very similar result arises from a 
mismatch between the $(CP)^2$ properties
of the Dirac and sterile mass eigenstates. 
We do not assume that the mass of the sterile component is
significantly larger than the mass of the Dirac components.

Following the same procedure as above, we have also calculated
$\mathcal{P}\left(\bar\nu_\mu\to\bar\nu_\mu\right)$,
$\mathcal{P}\left(\bar\nu_\mu\to\bar\nu_\tau\right)$,
$\mathcal{P}\left(\bar\nu_\mu\to\bar\nu_\zeta\right)$. These, together
with $\mathcal{P}\left(\bar\nu_\mu\to\bar\nu_e\right)$, sum to unity,
and carry no oscillatory term associated with the $(CP)^2 = -\openone$
sterile-neutrino component. Each of these contains a flavour
transmutation term. We have also calculated
$\mathcal{P}\left(\nu_\mu\to\nu_e\right)$ and find it equals
$\mathcal{P}\left(\bar\nu_\mu\to\bar\nu_e\right)$.

It is now to be noted that a non-zero $\chi$ mimics an anomalous
decay that produces a $\bar\nu_e$.  For this reason existing analysis
of the KARMEN and LSND data may be used to determine the value of
$\chi$.  Specifically, KARMEN saw $15$ $\bar\nu_e$ like
sequences. These are in agreement with the background expectation of
$15.8 \pm 0.5$ $\bar\nu_e$ like
sequences~\cite{Reichenbacher:2005nc}. If the LSND result is
assumed to be entirely due to an anomalous decay, then KARMEN should
have seen $15$ additional $\bar\nu_e$ like
events~\cite{Goldman:2007pc}. But because 
a non-zero $\chi$ mimics an anomalous decay, 
the  absence of the additional events at
KARMEN places severe constraints 
on $\chi$.  Therefore,
unless MiniBooNE reports 
a confirmation of a LSND-like signal 
with essentially
no  $L/E$ dependence, the existing data on KARMEN
rules out a non-zero $\chi$; i.e., it excludes a $(CP)^2 = -\openone$
sterile neutrino component
in the presented $3+1$ scenario.

The central result of this Letter establishes that the addition of
a
$(CP)^2 = - \openone$ sterile neutrino component does not introduce an
extra mass-squared difference in the neutrino oscillation probability.
It is, therefore, true that the observation of two independent
mass-squared differences does not necessarily fix the number of
underlying mass eigenstates as three. A $3+2$ extension of our formalism 
yields only $3$, not $4$, oscillatory terms in the relevant flavour oscillation
probabilities. 
For these reasons we argue that the 
$(CP)^2 = \pm \openone$ nature of any sterile neutrino component is 
experimentally accessible.

\begin{acknowledgments}
We wish to thank Terry Goldman,  Daniel Grumiller, and 
 Ben Leith for helpful discussions and comments.
We are grateful to an anonymous referee for bringing to
our attention the subject of nonorthogonal neutrinos~\cite{Langacker:1988up}.

\end{acknowledgments}
\bibliography{an}

\end{document}